\documentclass{article}
\usepackage{lineno}
\usepackage{xcolor}
\usepackage{amsmath}

\usepackage{PRIMEarxiv}

\usepackage[utf8]{inputenc} 
\usepackage[T1]{fontenc}    
\usepackage{hyperref}       
\usepackage{url}            
\usepackage{booktabs}       
\usepackage{amsfonts}       
\usepackage{nicefrac}       
\usepackage{microtype}      
\usepackage{lipsum}
\usepackage{fancyhdr}       
\usepackage{graphicx}       
\usepackage{caption}
\usepackage{subcaption}
\usepackage{float}

\begin{document}
\title{Analytical description of the time-over-threshold method based on the time properties of plastic scintillators equipped with silicon photomultipliers}

\author{
  N. Karpushkin, D. Finogeev, F. Guber, D. Lyapin, A. Makhnev, S. Morozov, D. Serebryakov \\
  Institute for Nuclear Research of the Russian Academy of Sciences \\
  60-letiya Oktyabrya prospekt 7a, Moscow 117312, Russia\\
  \texttt{karpushkin@inr.ru} \\
}

\maketitle

\footnote{\copyright 2024. This manuscript version is made available under the \href{https://creativecommons.org/licenses/by-nc-nd/4.0/}{CC-BY-NC-ND 4.0 license}. \\ Link to the formal publication: \href{https://doi.org/10.1016/j.nima.2024.169739}{https://doi.org/10.1016/j.nima.2024.169739}.}

\begin{abstract}
  A new highly granular neutron detector (HGND) for the identification and energy measurement of neutrons produced in nucleus-nucleus interactions at the BM@N experiment, Dubna, Russia, at energies up to 4 AGeV is under development. The detector consists of approximately 2000 fast plastic scintillators, each with dimensions of 40$\times$40$\times$25 mm$^3$, equipped with SiPM (Silicon Photomultiplier) with an active area of 6$\times$6 mm$^2$. The signal readout from these scintillators will employ a single-threshold multichannel Time-to-Digital Converter (TDC) to measure their response time and charge using the time-over-threshold (ToT) method. This article focuses on the analytical description of the signals from the plastic scintillator detectors equipped with silicon photomultipliers. This description is crucial for establishing the ToT to charge relationship and implementing slewing correction techniques to improve the time resolution of the detector.
\end{abstract}

\section{Introduction}
\label{sec:intro}
The BM@N experiment at NICA, JINR ~\cite{Afanasiev:2023opv, Kapishin:2020cwk, Senger:2022bzm}, aims to investigate nuclear matter properties at baryon densities 2-4 times higher than the nominal nuclear density, created in heavy-ion collisions with beam energies up to 4 AGeV. To measure neutron yields and azimuthal flows, providing valuable information about the contribution of the symmetry energy term to the equation of state of dense nuclear matter, a highly granular neutron detector (HGND) is currently in development~\cite{Guber_arxiv}.

The HGND consists of 16 alternating layers of scintillator and copper absorber, each with transverse dimensions of \(44 \times 44 \, \text{cm}^2\), all mounted on a common support frame. Each copper absorber layer is a single copper plate 3 cm thick. Each active scintillator layer consists of 121 scintillation cells measuring \(4 \times 4 \times 2.5 \, \text{cm}^3\), arranged in an \(11 \times 11\) matrix. The first scintillation layer serves as a veto detector, effectively distinguishing neutrons from charged particles. The total length of the HGND is approximately 1 meter, corresponding to about three nuclear interaction lengths.

The main structural element of the HGND is the scintillator cell. Various scintillator compositions were tested~\cite{Morozov_arxiv}. The JINR-produced scintillator, composed of polystyrene with the addition of 1.5\% paraterphenyl and 0.01\% POPOP, is considered the primary option. All surfaces of the cell, except for the one large face, are covered with a white diffuse reflector based on titanium dioxide TiO$_2$. The uncovered large face is oriented downstream relative to the beam. Its surface is polished and a silicon photomultiplier is centrally positioned on it using optical grease, while the remaining area of the surface is covered with black light-absorbing tape.

For scintillation signal detection, new fast silicon photomultipliers with epitaxial quenching resistors SiPM EQR15 11-6060D-S~\cite{NDL-EQR15} with a sensitive area of 6$\times$6 mm$^2$ are employed. In the remainder of the article, the sensitive element of the HGND, which consists of a JINR-produced scintillator cell equipped with a SiPM EQR15 11-6060D-S, is referred to as a scintillator detector.

Given that the HGND will have multiple readout channels $(\sim2000)$, the new FPGA based TDC (Field Programmable Gate Array based Time-to-Digital Converter) board with a 100 ps bin width has been developed~\cite{Finogeev_arxiv}. Each readout channel will connect to a comparator operating at a constant threshold. The HGND will record the response time relative to experiment trigger start time, enabling the determination of a particle's time-of-flight and subsequent kinetic energy calculation. The dynamic range of the scintillator detector spans approximately 0.5-7 MIP (Minimum Ionizing Particle). The detected charge will be indirectly measured by analyzing the time-over-threshold (ToT). To test the functionality of the HGND channels and align their time response, short-pulse light-emitting diodes attached to each scintillator cell will be used.

In this paper, an analytical model is proposed to describe an exponentially decaying light flash detected by a SiPM. The model allows for parameterizing the times of the leading and trailing edges at a constant threshold. This approach enables the introduction of necessary corrections to improve the time resolution and estimate the detected charge.

\section{Signals shapes from fast plastic scintillator and silicon photomultiplier}
\subsection{Processes inside the SiPM}
 The SiPM (Silicon Photomultiplier) comprises an array of SPAD (Single Photon Avalanche Diode) -- pixels. Upon absorption of a photon, an avalanche process is initiated within the corresponding pixel, leading to a rapid rise in current to its maximum value. The avalanche is subsequently quenched to ensure stable operation. In case of passive quenching this is accomplished through the incorporation of a quenching resistor. Classic SiPMs utilize polysilicon quenching resistors located on the device's surface, while one of the current development directions is focused on SiPMs with epitaxial quenching resistors~\cite{ZHAO2018252}. In such SiPMs, the quenching resistance is integrated within the epitaxial layer, increasing the density of cells. Reduced junction capacitance combined with a relatively low quenching resistance, leads to a fast recovery time. Additionally, the high geometric fill factor of these SiPMs allows for a wide dynamic range with high photon detection efficiency (PDE).
 
A comprehensive analysis of the processes in a SiPM, including consideration of all parasitic processes, can be found in the detailed study~\cite{Gundacker_2020}, while the pulse shape discussion is available in~\cite{Corsi_2006, Giustolisi_2013, DUARA2020164483}. In this article, an approximate description of the pulse shape is provided: the pulse response function of the SiPM is characterized by the time \(R_s C_T\), where \(R_s\) represents the sum of the load resistance and the low intrinsic SiPM resistance, and \(C_T\) is the total SiPM capacitance.

\subsection{Plastic scintillator flash readout with SiPM}
When charged particles traverse through a plastic scintillator, a flash of light is emitted, yielding a measurable signal. Its form can be parameterized as given in Equation~\ref{eq:light_sc}. Here, $N_{ph}^0$ is the normalization factor, and the rise time $\tau_R$ and decay time $\tau_D$ define the time characteristics of the scintillation flash. Notably, the rise time is typically negligible, allowing its omission in subsequent analyses. The number of photons produced in scintillator at a time $t$ is therefore given by:

\begin{equation}
  \label{eq:light_sc}
  N^{scint}_{ph}(t) = N_{ph}^0 (1-e^{-t/\tau_R})e^{-t/\tau_D} \approx N_{ph}^0 e^{-t/\tau_D}.
\end{equation}

The generated discharging photoelectron current in this case can be parameterized as
\begin{equation}
  \label{eq:IpeScint}
  I_{discharge} = -q \eta G \frac{dN^{scint}_{ph}}{dt} = \frac{\eta q G N_{ph}^0}{\tau_D}  e^{-t/\tau_D},
\end{equation}
where $\eta$ represents the photon detection efficiency, $G$ is the SiPM gain, and $q$ represents the absolute elementary charge.

The voltage across the load resistance can be obtained as a product of the load resistance and the convolution of the discharging photoelectron current \(I_{discharge}\), with the pulse response function of the SiPM, denoted as \(g(t) = \frac{1}{R_s C_T} e^{-t/R_s C_T}\). The pulse response function is normalized to unity. 

\begin{equation}
  \label{eq:V_conv}
  V(t) = R_s (I_{discharge}*g)(t) = \frac{\eta q G N_{ph}^0 }{C_T \tau_D} \int_0^t e^{-\frac{x}{\tau_D}} e^{-\frac{t-x}{R_s C_T}} dx = \frac{ \eta q G N_{ph}^0 R_s}{R_s C_T - \tau_D} \left(e^{-t/R_s C_T} - e^{-t/\tau_D}\right) .
\end{equation}

A similar solution can be obtained from the differential equation if the SiPM is considered "integrally". The current through the SiPM is determined by two simultaneous processes: the discharging current and the recharge current.The recharge current is determined at each moment of time only by the value of the charge \(Q\) stored in the SiPM.

\begin{equation}
  \frac{dQ}{dt} = I_{recharge} - I_{discharge} \quad \quad V_{bias} - R_s I_{recharge} = \frac{Q}{C_T}.
\end{equation}

Solving such a differential equation with the initial condition \(Q(0) = C_T V_{bias}\), one obtains an expression for the voltage across the load resistor:

\begin{equation}
  \label{eq:V_diffur}
  V(t) = R_s I_{recharge} = \frac{\eta q G N_{ph}^0 R_s}{R_s C_T - \tau_D} \left(e^{-t/R_s C_T} - e^{-t/\tau_D}\right).
\end{equation}

\subsection{Addressing derivative discontinuity}
The solution for the voltage across the load resistance (Equations~\ref{eq:V_conv} and~\ref{eq:V_diffur}) assumes equal amplitudes of two exponential terms. However, in reality, these terms exhibit slightly different amplitudes; specifically, the one with the smaller time constant (in our case $\tau_D < R_s C_T$) appears to have a slightly smaller amplitude. Information about the time constants and amplitudes of exponential terms is derived from spectral analysis using the Prony Least Squares method~\cite{Prony_2019}, and is discussed in section~\ref{sec:comparison}. This amplitude difference is parameterized by a factor $(1-p),$ which can be interpreted as a phase shift: $ (1-p)e^{-t/\tau_D} = e^{\frac{-t + \tau_D \ln(1-p)}{\tau_D}} $. Consequently, the term responsible for discharge leads the component related to recharge.

Additionally, the solution (Equations~\ref{eq:V_conv},~\ref{eq:V_diffur}) is valid for time $t \geq 0$ and exhibits a sharp start at zero. In reality, the baseline smoothly transitions into the signal due to the limited passband of the SiPM, transmission line, and other electronics. Therefore, the obtained solution must be modified to account for this. To achieve a more gradual transition at the beginning of the signal, a smoothly increasing parameterization in that region is incorporated. The solution (Equations~\ref{eq:V_conv},~\ref{eq:V_diffur}) is shifted in time by a value of $T_p,$ which is the time when the signal reaches a certain constant fraction of its amplitude \(pV_0\). The conditions of the resulting solution are then matched with a smooth solution at this point in time $T_p$.

\[
  V_0 \equiv \frac{\eta q G N_{ph}^0 R_s}{R_s C_T - \tau_D} \quad \quad T_p = \frac{2e-1}{e-1} \frac{p R_s C_T \tau_D}{R_s C_T(1-p) - \tau_D}
\]

\begin{equation}
  \label{eq:model}
  V(t) = 
  \begin{cases}
    \frac{pV_0}{e-1}\frac{t}{T_p}\left(e^{\frac{t}{T_p}}-1\right) & \text{if } 0 \leq t < T_p \\
    V_0\left(e^{-\frac{t-T_p}{R_s C_T}}-(1-p)e^{-\frac{t-T_p}{\tau_D}}\right) & \text{if } t \geq T_p
  \end{cases}
\end{equation}

\subsection{Comparison of model with experimental data}
\label{seq:Scintillator_Data}
\subsubsection{Experimental setup}
\begin{figure}[htbp]
  \centering
  \includegraphics[width=.7\textwidth]{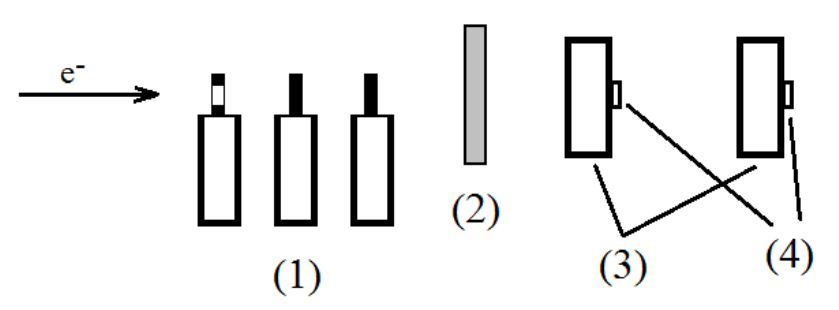}
  \caption{Schematic view of the setup for measurements of scintillator characteristics using the electron beam of the ''Pakhra'' synchrotron (LPI, Troitsk). (1) Beam counters, (2) Lead bar, (3) Scintillator cells, (4) SiPMs.}
  \label{fig:setup}
\end{figure}

The data from scintillator cells was acquired using a secondary electron beam with an energy of 280 MeV at the ''Pakhra'' synchrotron at LPI~\cite{Alekseev_2019}. The experimental setup is shown in Figure~\ref{fig:setup}. A veto counter with a hole diameter of 10 mm, followed by two scintillation beam detectors with transverse dimensions of \(10 \times 10 \, \text{mm}^2\) and a thickness of 5 mm, were placed on the beam axis and used in the trigger (1). Subsequently, lead bar (2) and two scintillator cells (3) with SiPMs (4), were positioned further along the beam. For each scintillator detector under study, the analog signal and signal from the comparator were digitized using a high-precision 12-bit ADC (CAEN DT5742) with a sampling rate of \(5 \, \text{GS/s}\).

A similar setup was used in~\cite{Morozov_arxiv}, resulting in a scintillator detector time resolution of 115 ps with slewing correction applied by the signal amplitude and accounting for the start counter's intrinsic time resolution.

This paper is focused on establishing the ToT to charge relationship to compensate for slewing effects using the ToT measured by the comparator. For this purpose, 17 pairs of scintillator cells (3) were measured using the setup provided in Figure~\ref{fig:setup}. During these measurements, a 1 cm thick lead bar (2) was introduced into the electron beam in front of the scintillators under test to increase the amplitude spread and make the experimental conditions more consistent with those of the future full-scale detector. The comparator time was analyzed relative to the time of the analog signal at a constant fraction (20\%) of its amplitude, calculated from the digitized data. Further analysis is illustrated for one of the 34 tested scintillators.

\subsubsection{Comparison with model}
\label{sec:comparison}
\begin{figure}[htbp]
  \centering
  \begin{subfigure}[htbp]{\textwidth}
    \centering
    \includegraphics[width=0.5\textwidth]{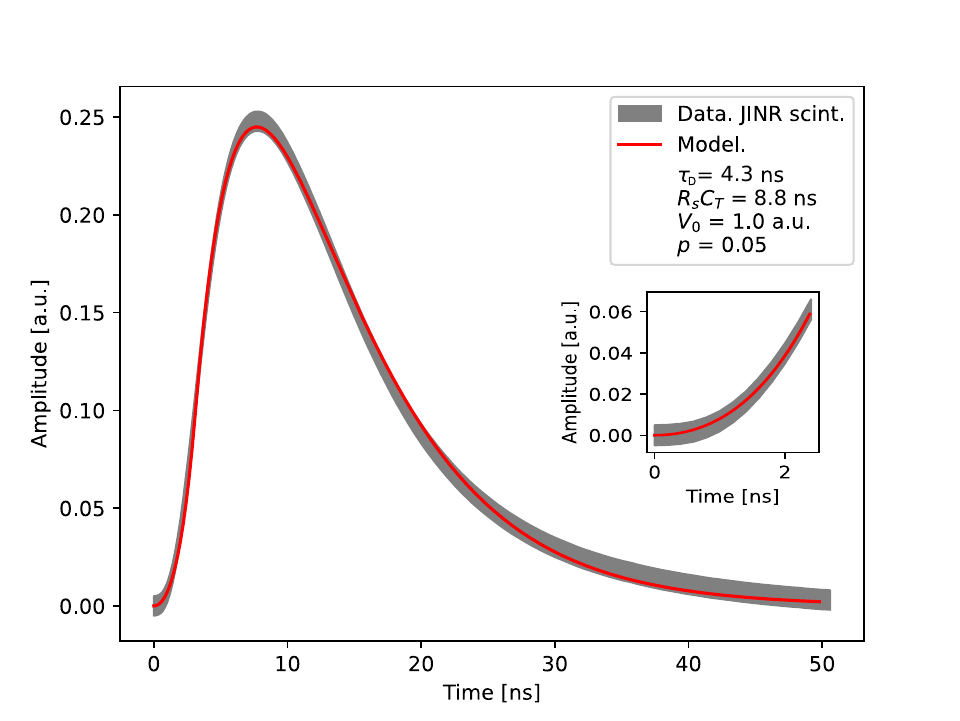}
    \caption{}
    \label{fig:wfm}
  \end{subfigure}
  \begin{subfigure}[htbp]{0.45\textwidth}
    \centering
    \includegraphics[width=\textwidth]{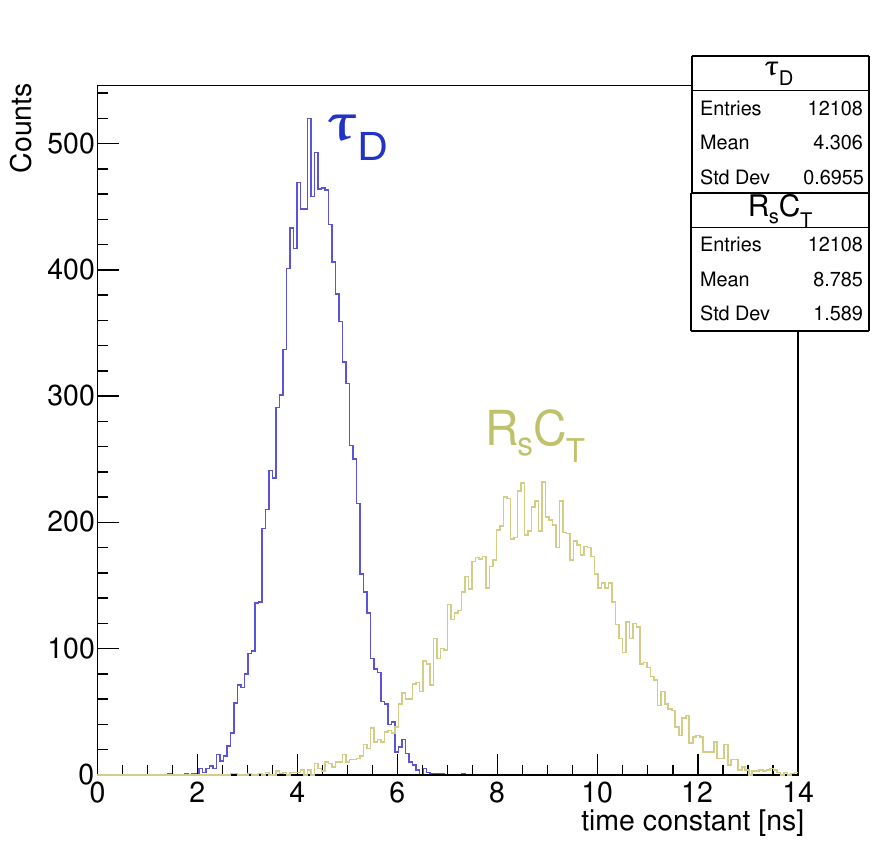}
    \caption{}
    \label{fig:taus}
  \end{subfigure}
  \begin{subfigure}[htbp]{0.46\textwidth}
    \centering
    \includegraphics[width=\textwidth]{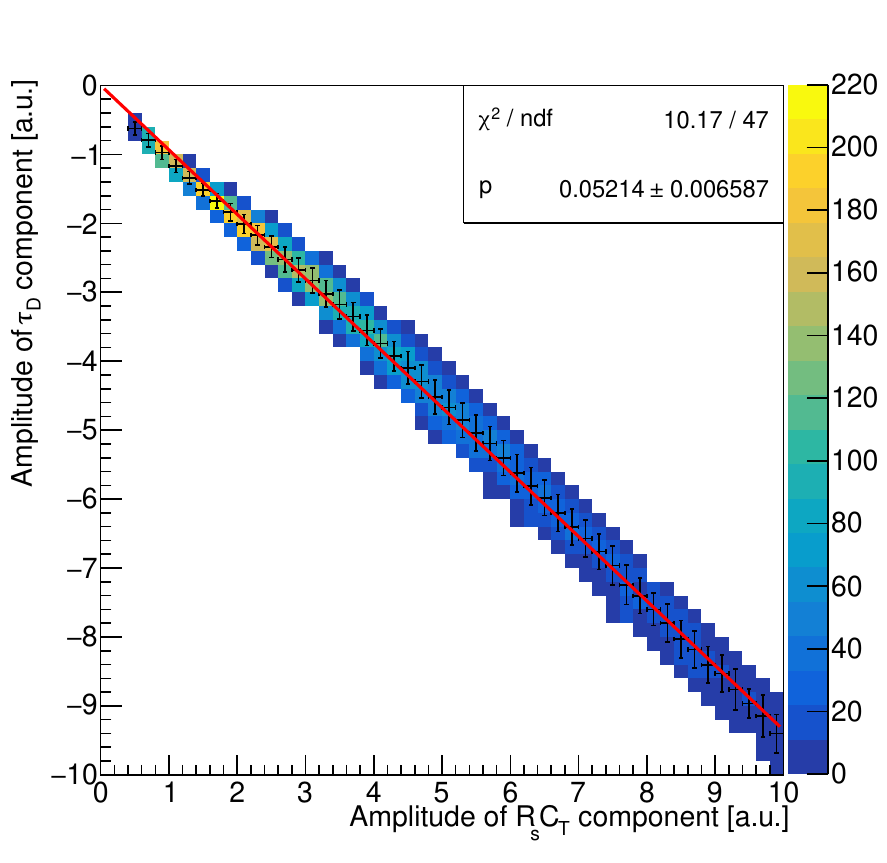}
    \caption{}
    \label{fig:ampls}
  \end{subfigure}
  \caption{(a) Comparison between the measured data curve and model parametrisation. The inset illustrates the enlarged timescale at the beginning of the signal. (b) Distributions of the characteristic discharge (blue) and recharge (yellow) times of the parametrization. (c) Correlation of corresponding amplitudes. The profile (black) indicates the standard deviation error bars and is fitted with a linear function. The data were acquired using scintillator detector during tests at the 280 MeV electron beam.}
  \label{fig:wfm_sc}
\end{figure}

The parametrization given in Equation~\ref{eq:model} is utilized to reproduce the signal from scintillator detector, and is illustrated in the Figure~\ref{fig:wfm}. 

The dark grey area represents multiple waveforms superimposed and normalized to present the characteristic data curve. The inset illustrates the enlarged timescale at the beginning of the signal. The red line corresponds to a model (Equation~\ref{eq:model}) with parameters $\tau_D, R_s C_T,$ and $p,$ derived from spectral analysis using the Prony Least Squares method~\cite{Prony_2019}. In this method, each waveform is decomposed into exponential components, and the values of their time constants and amplitudes are estimated. Distributions of both characteristic discharge and recharge times $\tau_D$ and $R_s C_T$ are provided in Figure~\ref{fig:taus} while correlation of their amplitudes ($-V_0(1-p)$ vs $V_0$) with linear fit is given in Figure~\ref{fig:ampls}.

The theoretical characteristic recharge time, determined by a load resistance of $R_l = 33 \Omega$ and a SiPM capacitance of $C_T \approx 200$ pF~\cite{NDL-EQR15}, is approximately 6.7 ns. However, the observed recharge time in the data is $8.8 \pm 1.6$ ns, indicating the presence of additional intrinsic resistance within the SiPM. The amplitudes of the exponential components exhibit a linear correlation. The slope, parameterized as $-(1-p)$ and derived from the linear fit, indicates that the value of $p$ is approximately 0.05. 

\section{Slewing correction and ToT method}
\label{sec::slewing}
\subsection{Slewing correction}
The key information to extract from the data is the response time relative to the trigger time, as it determines the particle's time-of-flight and enables the calculation of its kinetic energy. However, when measuring time, a known issue arises regarding the dependency of the threshold crossing time on the signal amplitude, commonly referred to as time walk or slewing. Given the known waveform parameterization from Equation~\ref{eq:model}, straightforward calculations yield the following expressions for identifying the times at which the leading edge \((t_1)\) and the falling edge \((t_2)\) cross the threshold $\theta$:

\begin{equation}
  \label{eq:times}
  t_1 \approx T_p \sqrt{\frac{\theta(e-1)}{pV_0}}
  \quad \quad
  t_2 \approx T_p + R_sC_T \ln \frac{V_0}{\theta}.
\end{equation}
In the derivation of the first equation, the Taylor series expansion around \( t \to 0 \) is utilized. In the second equation, the term responsible for the SiPM discharge is neglected. This is justified as the impact of this term, with its small time constant, becomes negligible compared to the term responsible for the recharge at a larger time scale.

To address slewing correction, one can approximate it by calculating the Time-over-Threshold (ToT). By expressing the \(V_0/\theta\) relationship from the right side of Equation~\ref{eq:times}, substituting it into the left side of Equation~\ref{eq:times}, and replacing \(t_2\) introducing the definition of ToT, one obtains the following parameterization of \(t_1\) as a function of ToT:

\begin{equation}
  \label{eq:timewalk}
  t_1 \approx 2 R_sC_T W_0 \left( \frac{T_p}{2R_sC_T} \sqrt{\frac{e-1}{p}} e^{\frac{T_p - ToT}{2R_sC_T}} \right) \quad \quad ToT \equiv t_2 - t_1,
\end{equation}
where \(W_0\) denotes the principal branch of the Lambert W function~\cite{Lambert_1996}.

\begin{figure}[htbp]
  \centering
  \includegraphics[width=.6\textwidth]{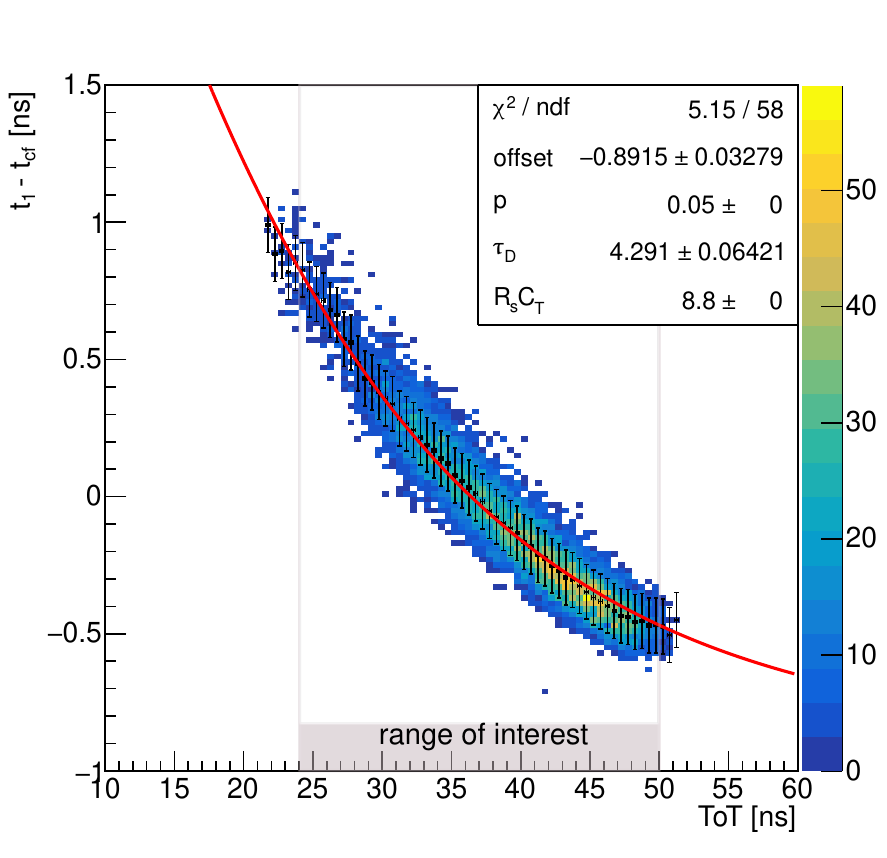}
  \caption{Correlation of the leading edge time \(t_1\), measured relative to constant fraction time \(t_{cf}\), and ToT. Data were obtained from one of the tested scintillator detector using a 280-MeV electron beam. The black profile indicates the error bars and is fitted with the model given by Equation~\ref{eq:timewalk}.}
  \label{fig:tdc_tot}
\end{figure}

Figure~\ref{fig:tdc_tot} illustrates the correlation between the leading edge time of the signal (\(t_1\)) and its ToT, measured using the setup described in Section~\ref{seq:Scintillator_Data}. The black profile shows the error bars corresponding to the time resolution obtained in~\cite{Morozov_arxiv}. The \(t_1\) time is measured by a comparator and is plotted relative to the \(t_{cf}\) (constant fraction time) when the analog signal reaches 20\% of its amplitude, calculated from the digitized data. This approach ensures that the slewing is reflected solely in \(t_1\). The \(t_{cf}\) time determined by the signal crossing a constant fraction of its amplitude is negligibly susceptible to slewing and affects only the offset parameter, which is of no significance.

The red line in the Figure~\ref{fig:tdc_tot} represents a parametrization described by Equation~\ref{eq:timewalk}. It was determined that this dependence has low sensitivity to parameters \(p\) and \(R_sC_T\). Therefore, these parameters were fixed at their average values extracted from the spectral analysis described in section~\ref{sec:comparison}, while parameter \(\tau_D\) was left free to account for fluctuations in the properties of the scintillators. The parameter values are provided in the figure's inset, while the range of interest indicates the scintillator detector's dynamic range. The parametrization given in Equation~\ref{eq:timewalk} effectively describes the data within this range.

The time resolution of the scintillator detectors was determined by dividing by \(\sqrt{2}\) the width of the distribution of the time difference between two scintillator detectors under study. Slewing correction was applied to each scintillator detector according to Equation~\ref{eq:timewalk}. The resulting average time resolution over the 34 tested scintillator detectors is approximately 130 ps. This is an improvement of about 50\% compared to the time resolution without slewing correction and is approximately 15 ps worse compared to the time resolution obtained for this type of scintillator with slewing correction by the signal amplitude~\cite{Morozov_arxiv}.

\subsection{Charge estimation from ToT}
Reconstruction of the detected charge from the measured ToT allows for the determination of reconstruction uncertainties. The total charge \(Q_{total}\) transferred through the load resistor, can be derived from Equation~\ref{eq:model} as the area under the curve divided by the load resistance:

\begin{equation}
  \label{eq:amplitude}
  Q_{total} = \frac{V_0}{R_s} \left(\frac{pT_p}{2(e-1)} + R_sC_T - \tau_D(1-p)\right) \equiv \frac{V_0\lambda}{R_s} ,
\end{equation}
\noindent where \(\lambda\) is a redesignation of the time constant to simplify further explanation.

The \(Q_{total}\) as well as amplitude has a linear relationship with the normalisation parameter \(V_0\) and therefore has the same resolution. Following Equations~\ref{eq:times} and~\ref{eq:timewalk}, one gets:

\begin{equation}
  \label{eq:qdc_tot}
  Q_{total} = \frac{\theta \lambda}{R_s} e^{\frac{ToT + t_1 - T_p}{R_sC_T}}.
\end{equation}

\begin{figure}[htbp]
\centering
\includegraphics[width=.6\textwidth]{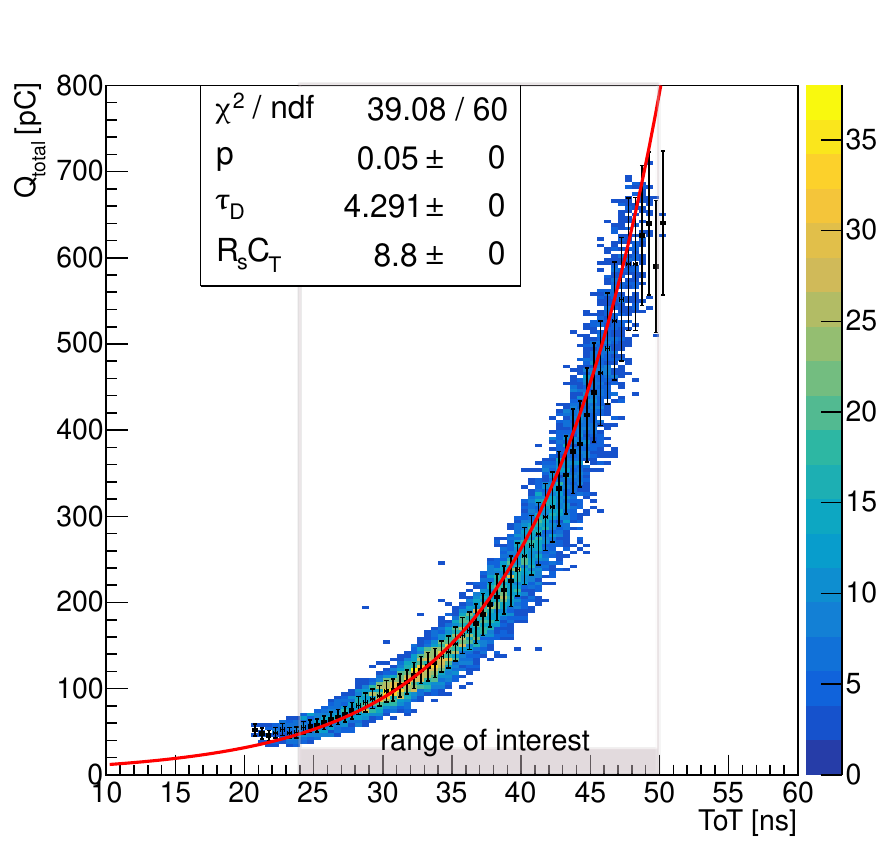}
\caption{Correlation of \(Q_{total}\) and ToT. Data were obtained from one of the tested scintillator detector using a 280-MeV electron beam. The black profile indicates the error bars and is fitted with the model given by Equation~\ref{eq:qdc_tot} with fixed parameters.}
\label{fig:qdc_tot}
\end{figure}

The \(Q_{total}\) in the data can be determined numerically by calculating the area under the curve shown in Figure~\ref{fig:wfm} and dividing it by the load resistance \(R_s\). To estimate the charge resolution, the correlation between ToT and \(Q_{total}\) is utilized. This correlation was measured with the setup described in Section~\ref{seq:Scintillator_Data} for all 34 scintillator detectors. Correlation for one of them is displayed in Figure~\ref{fig:qdc_tot}. The black profile displays error bars where the x-axis error represents the time resolution obtained in~\cite{Morozov_arxiv}, and the y-axis error denotes the standard deviation of the charge distribution corresponding to the specific ToT bin. The red line in the figure represents the parametrization of this dependence using Equation~\ref{eq:qdc_tot} with fixed parameters. The values of \(p\) and \(R_sC_T\) were fixed at their average values, while the parameter \(\tau_D\) was obtained from the fit shown in Figure~\ref{fig:tdc_tot}. The parametrization effectively describes the data within the range of interest.

To estimate the measurement and model uncertainties, we scanned over narrow ToT bins. The measurement uncertainty was determined by examining the width of the charge distribution within each ToT bin. The model uncertainty was assessed by comparing the mean value of the charge distribution to the analytical parameterization. The values of measurement and model relative errors obtained in this manner do not exceed 15\% and 10\%, respectively, for all tested scintillator detectors within the range of interest. The deviation of the parameterization from the \(Q_{total} : ToT\) dependence profile at high \(Q_{total}\) is attributed only to the limited available statistics in this region and is not considered in the resolution analysis.

\section{Summary}
An analytical description of the signals from fast scintillators with SiPM readout, focusing on slewing correction by ToT, has been proposed for the highly granular neutron detector of the BM@N experiment. The application of the slewing correction technique improves the time resolution by approximately 50\%, achieving a resolution of 130 ps for the JINR-produced scintillator equipped with the SiPM EQR15 11-6060D-S, meeting the requirements of the detector project~\cite{Guber_arxiv}. The time resolution obtained with the slewing correction by ToT is only $\sim15$ ps worse compared to the time resolution achieved with slewing correction based on the measured signal amplitude. This result is considered satisfactory and represents an acceptable compromise for the opportunity to simplify the design and to reduce the cost of the acquisition system by using only TDCs. The designed model effectively describes the data within the region of interest for all 34 tested scintillator detectors and provides a method to estimate the signal charge from its ToT, resulting in a charge resolution better than 20\%.

\section*{Acknowledgments}
This work was carried out at the Institute for Nuclear Research, Russian Academy of Sciences, and supported by the Russian Scientific Foundation grant \textnumero 22-12-00132.

\bibliographystyle{elsarticle-num} 

\end{document}